\documentclass[aps,prl,reprint,groupedaddress,twocolumn]{revtex4}
\usepackage{graphicx}   
\usepackage{amsmath,amssymb,wasysym}
\usepackage[amsmath,thmmarks,framed]{ntheorem}
\usepackage{color}
\usepackage{hyperref}

\begin{document}

\newcommand{\idunno}[0]{\fcolorbox{red}{blue}{\color{white} 666}}
\newcommand{\lolwut}[1]{\fcolorbox{red}{blue}{\color{white} #1}}
\newcommand{\cesium}[0]{$^{133}$Cs}
\newcommand{\uW}[0]{\,$\mu$W}
\newcommand{\fthz}[0]{\,fT/$\sqrt{\text{Hz}}$}
\newcommand{\pthz}[0]{\,pT/$\sqrt{\text{Hz}}$}
\newcommand{\brf}[0]{$\vec{B}_{\text{rf}}$}
\newcommand{\brfm}[0]{\vec{B}_{\text{rf}}}
\newcommand{\bmain}[0]{$\vec{B}_0$}

\newcommand{\labnoiseSTEP}[0]{40\fthz}
\newcommand{\labnoiseSTEPLS}[0]{55\fthz}
\newcommand{\labnoiseLSD}[0]{40\fthz}
\newcommand{\shotnoise}[0]{1.7\fthz}
\newcommand{\SNLKitching}[0]{2\pthz}
\newcommand{\pumpnoise}[0]{3.3\fthz}
\newcommand{\shotnoiseSync}[0]{1.9\fthz}


\title{Alkali-vapor magnetic resonance driven by fictitious radiofrequency fields}


\author{Elena Zhivun}
\affiliation{Department of Physics, University of California, Berkeley, California 94720-7300, USA}
\author{Arne Wickenbrock}
\affiliation{Institut f\"{u}r Physik, Johannes Gutenberg Universit\"{a}t, 55128 Mainz, Germany}
\email[]{wickenbr@uni-mainz.de}
\author{Brian Patton\footnote{Present address: AOSense, Inc., Sunnyvale, CA 94085,
USA.}}
\affiliation{Department of Physics, University of California, Berkeley, CA 94720-7300 and\\
Physik-Department, Technische Universit\"{a}t M\"{u}nchen, 85748 Garching, Germany}
\author{Dmitry Budker}
\affiliation{Helmholtz Institut, Johannes Gutenberg Universit\"{a}t, 55099 Mainz, Germany\\
Department of Physics, University of California, Berkeley, CA 94720-7300 and\\ Nuclear Science Division, Lawrence Berkeley National Laboratory, Berkeley, CA 94720
}

\author{Dedicated to Professor William Happer on the occasion of his 75th birthday}



\date{\today}

\begin{abstract}
We demonstrate an all-optical \cesium{} scalar magnetometer, operating in nonzero magnetic field, in which the magnetic resonance is driven by an effective oscillating magnetic field provided by the AC Stark shift of an intensity-modulated laser beam. We achieve a projected shot-noise-limited sensitivity of \shotnoise{} and measure a technical noise floor of \labnoiseSTEP{}. These results are essentially identical to a coil-driven scalar magnetometer using the same setup. This all-optical scheme offers advantages over traditional coil-driven magnetometers for use in arrays and in magnetically sensitive fundamental physics experiments e.g., searches for a permanent electric dipole moment of the neutron.
\end{abstract}


\maketitle



A far-off-resonant laser beam can affect the energy levels of an atom even in the absence of absorption. These shifts can be decomposed into scalar, vector, and tensor contributions \cite{Happer68,LS1972,DimaOpticalPolarizedAtoms}. While the scalar component shifts the Zeeman sublevels together and forms the basis for optical trapping experiments, the vector light shift acts differently on individual magnetic sublevels. It influences the energy structure like the Zeeman effect; the vector light shift of a circularly polarized laser beam acts like a fictitious magnetic field directed along the propagation axis. The strength of this fictitious field is proportional to the beam intensity and the atomic vector polarizability. 
In this work we focus on effects of the vector light shift (later referred to as only ``light shift" or LS). The phenomenon has been used to manipulate atomic spins all-optically, e.g. driving magnetic resonance \cite{LS1972,Na1990,mr1995,Happer67}, inducing spin echoes \cite{echo1990,echo1991,echo2008}, enabling phase control \cite{LS2009}, deflecting atoms \cite{grad2002} and using cold atoms for precise polarization measurement \cite{LS2013}. A modulated fictitious magnetic field has also recently been used to help scan through the zero-field condition in a dc magnetometer with \SNLKitching{} sensitivity \cite{Kitching2014} and to convert a scalar into a vector magnetometer \cite{Lena2014}. In this paper, oscillating fictitious magnetic fields are the key component of an all-optical finite-field rf-driven magnetometer.


Optically pumped rf-driven magnetometers have been widely used for precise measurements since the 1950s \cite{BellBloom}.
They have a broad range of applications and are themselves still the subject of research \cite{MX2014}. 
The general working principle relies on optically polarizing the atoms along the leading magnetic field \bmain{} and then driving transitions between magnetic sublevels via a resonant radiofrequency magnetic field \brf{}.
Arrays of rf-driven magnetometers are employed when monitoring the magnetic field in an area or volume is required. 
This includes medical applications such as human heart or brain activity mapping \cite{Array1,Array2}, or in fundamental-physics experiments e.g., those searching for a permanent electric dipole moment of the neutron (nEDM) \cite{edm2005,edm2006}. 
In these applications, the rf technique has important limitations. The oscillating magnetic field contaminates the monitored environment which is detrimental for precise measurements. Additionally, crosstalk between adjacent sensors places a limit on spatial resolution of a sensor array. 
We demonstrate a way to overcome these limitations by replacing the radio frequency coils with intensity-modulated laser beams. All-optical rf-driven magnetometry could also be useful in remote magnetometry applications, where real magnetic fields cannot be directly applied to the atomic sample \cite{Higbie2011,Patton2012}.



%
\begin{figure}[!htb]
\centering
\includegraphics{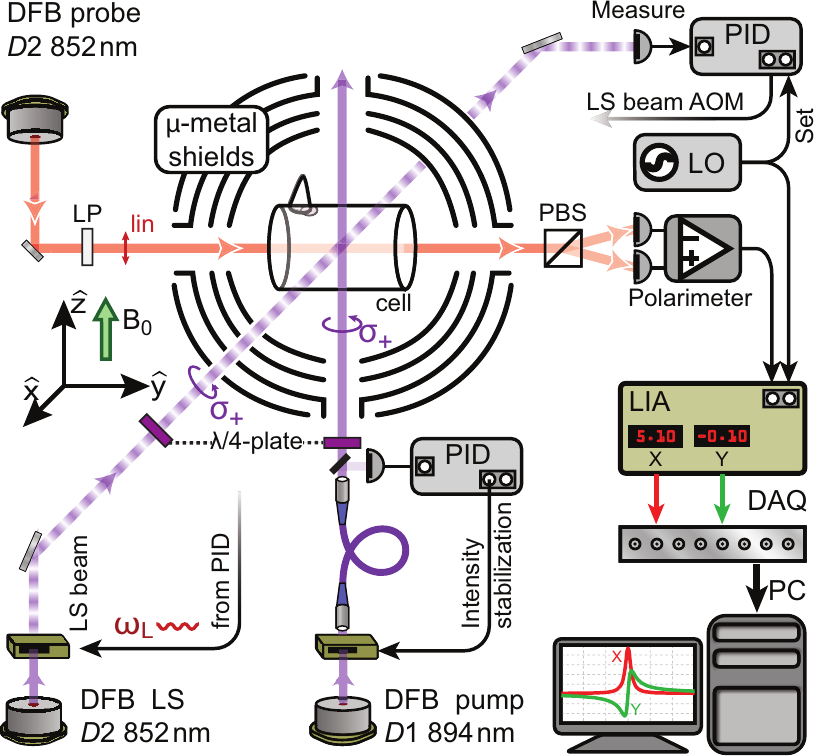}
\caption{Experimental setup. The vapor cell is housed in a four-layer {$\mu$-metal} shield. An unmodulated intensity-stabilized pump beam is circularly polarized and propagates along the leading field in the \^{z} direction. Spin precession is induced either by an oscillating magnetic field \brf{}$\parallel$\^{x} or by the modulated light-shift (LS) beam with $\text{\^{k}}_\text{LS}\!\!\parallel\!\!\text{\^{x}}$. A local oscillator (LO) provides the control signal for either the rf coil or the LS acousto-optic modulator (AOM), corresponding to the method used to induce the spin precession. A linear polarizer (LP) ensures a linearly polarized probe with $\text{\^{k}}_\text{pr}\!\!\parallel\!\!\text{\^{y}}$. The beam experiences rotation of its polarization plane, which is detected by a polarimeter, consisting of a polarizing beam splitter (PBS) and two photodiodes. The precession signal is demodulated by a lock-in amplifier (LIA) with the LO as a reference, and digitized by the PC via a data acquisition board (DAQ). All light sources used are distributed feedback laser (DFB).}
\label{fig:setup}
\end{figure}

Our rf-driven magnetometer has a projected shot-noise-limited sensitivity of \shotnoise{} which brings it within an order of magnitude to the most sensitive magnetometers \cite{mag054fT,mag7fT,mag9fT} in the finite-field regime.
Its experimentally demonstrated sensitivity is \labnoiseSTEP{} as measured in the laboratory, 
where technical noise limits the stability of the leading field.
Replacing the radiofrequency coil with an intensity-modulated laser beam reproduces the sensitivity and only slightly increases the noise floor of the rf-driven magnetometer at frequencies below 1~Hz. Analysis of the power spectrum of the magnetic-field changes shows that both magnetometers reach the same noise floor around 1.5~Hz and that the increased root mean square noise of the light-shift magnetometer is mostly due to low-frequency components. Overall, this demonstrates a direct improvement to current experiments using arrays of rf-driven magnetometers.


%


The experimental setup is shown in Fig. \ref{fig:setup}. An evacuated, paraffin-coated \cesium{} cell is placed within a four-layer \mbox{$\mu$-metal} magnetic shield with a total attenuation factor of $\sim\!\!10^{-6}$ \cite{DimaOpticalMagnetometry}.  
The cell is a 50\,mm long cylinder with a diameter of 50\,mm. 
Within the cell, the longitudinal spin relaxation time of the \cesium{} vapor is measured to be $T_1=0.7$\,s (here $T_1$ is the longer relaxation time in the bi-exponential decay as discussed in \cite{GrafDecay}). 

The temperature of the cell is stabilized around 17.5\,$^{\circ}\text{C}$ by a flow of cooled air. 
Coils inside the shields provide the leading field \bmain{}, gradient compensation along the \^{z} direction, and an oscillating \brf{} field in the conventional rf-driven magnetometer setup.
The leading-field and the gradient compensation coils are powered with separate channels of a custom current source (Magnicon GmbH) that exhibits a stability of $\sim\!\!10^{-7}$ over 100 seconds. The leading field was $\sim\!\!480$\,nT corresponding to a Larmor frequency of 1690\,Hz.


The pump beam (894\,nm DFB laser, 75.5\,$\mu$W) is launched parallel to \bmain{} ($\text{\^{k}}_\text{pu}\!\!\parallel\!\!\text{\^{z}}$) by a polarizing fiber (Fibercore HB830Z). 
The light power at the fiber output is stabilized by an AOM in a feedback loop, thus minimizing the fluctuations of the light shift induced by the pump beam. A zero-order quarter-wave plate before the magnetic shield ensures circularly polarized light.
The pump frequency is locked to the $D1$ $F = 3 \rightarrow F' = 4$ transition by a dichroic atomic vapor laser lock (DAVLL) \cite{DAVLL1,DAVLL_Budker,DAVLL2}.
Since we observe a magnetic resonance within the $F=4$ manifold, the probe is tuned to the $D2$ $F = 4 \rightarrow F' = 5$ transition, stabilized by a seperate DAVLL. The pump polarizes the $F=4$ manifold by depopulating the $F=3$ manifold,
while producing minimal light shift and broadening of the $F=4$ magnetic resonance \cite{Szymek}.
Moreover, we can see narrowing ($\sim$ 10\%) of the spin-exchange-limited resonance line \cite{LightNarrowingHapper,Schulte2011} due to the high polarization in the vapor.

The $\sigma+$ polarized light shift beam (852\,nm DFB laser, 1.9\,mW time-averaged power equivalent to a fictitious magnetic field with 0.19\,nT amplitude)  propagates orthogonally to \bmain{} with $\text{\^{k}}_\text{LS}\!\!\parallel\!\!\text{\^{x}}$. The intensity of the laser is modulated by an AOM in order to provide the time-varying light shift.
For this, we stabilize the transmission of the light-shift beam with a feedback loop and then modulate the set-point with a sinusoidal control signal of a function generator. This way we make sure to drive the resonance with a single harmonic signal independent of any nonlinear electronic response in the circuit.
The optical frequency of this laser is detuned 50\,GHz below the $D2$ $F = 4 \rightarrow F' = 5$ transition and frequency locked via a wavemeter (\r{A}ngstrom/HighFinesse \mbox{WS-7}). 
This detuning was chosen to minimize optical pumping by the light-shift beam while still having sufficient fictitious field to drive the magnetic resonance. 




We note here that the laser beams do not overlap, and their waist sizes ($<1$\,mm) are small compared to the cell diameter (50\,mm). Since the atoms traverse the cell in a time shorter than the precession period, they motionally average the magnetic fields (real and fictitious) within the cell, as well as the intensity of the pump and probe beams \cite{SpatialAveraging2006}. For this reason the observed light shifts only depend upon the power of the beams and not their spatial intensity profiles.

\begin{figure}[tbh]
	\centering
\includegraphics{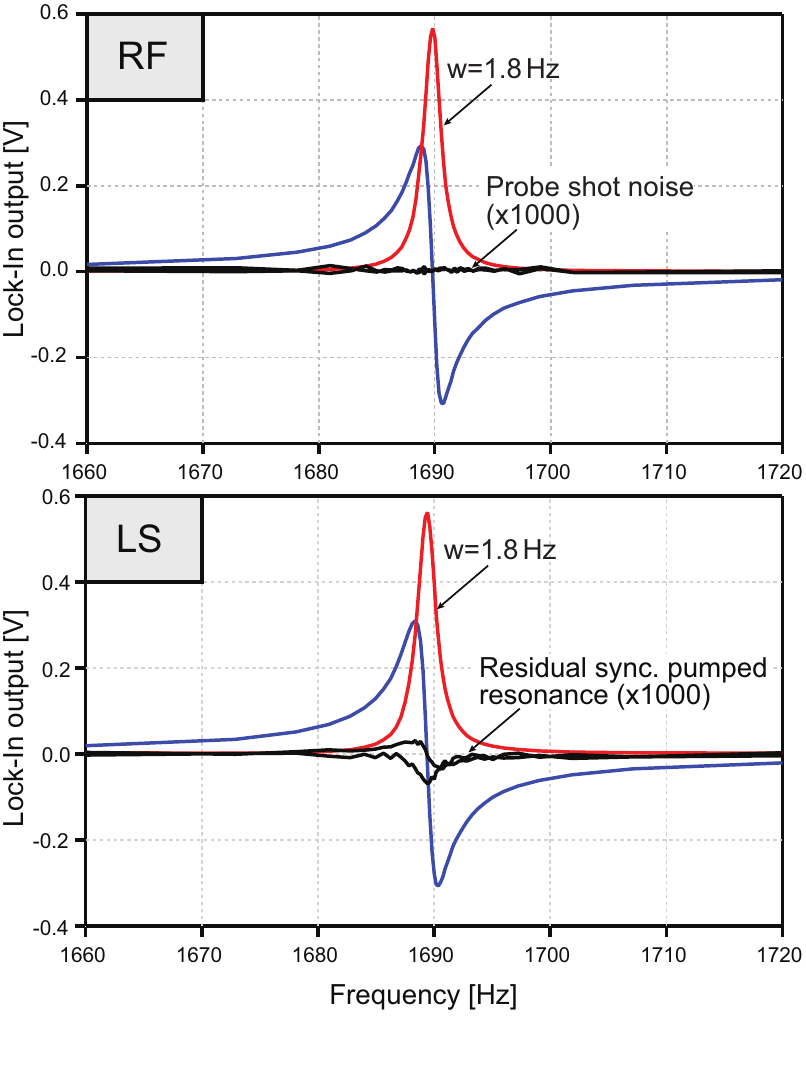}
	\caption{Driven-oscillation scans. 
	Top: the quadrature outputs of the lock-in amplifier for rf-driven magnetometer. The blue and red curves are the dispersive (\textsf{Y}) and the absorptive (\textsf{X}) signals. In black, magnified by a factor of 1000, is the same scan with the pump beam blocked. Bottom: the same resonance for the LS mode. The curve exhibits the same width as the coil-driven magnetometer for the same amplitude. The small residual resonance seen in the lower plot is due to synchronous pumping by the far-detuned LS beam.}
	\label{fig:Figure2}
\end{figure}

We compare two modes of magnetometer operation. 
The first mode is an rf-driven magnetometer (RF), where the spin precession is induced by an oscillating magnetic field \brf{}$\parallel$\^{x}. 
The second mode is a light-shift magnetometer (LS), where an oscillating fictitious magnetic field in the same direction is induced by an intensity-modulated laser beam. 
In both cases the sinusoidal driving signal is produced by a frequency synthesizer (Berkeley Nucleonics Corp. BNC645) that plays the role of the local oscillator (LO in Fig. \ref{fig:setup}). 
The precession of the \cesium{} magnetization is read out by a balanced polarimeter that detects the probe polarization rotation. The polarimeter's electronic noise corresponds to shot noise of 0.45\,$\mu$W of light (0.17\,\fthz{} equivalent magnetic field noise). 
The probe beam (852\,nm DFB laser, 16.4\,\uW{}) propagates along \^{y} and its linear polarization is orthogonal to $\vec{B}_0$ in order to minimize probe-induced static atomic alignment (quadrupole moment, see \cite{DimaAlignment}). 
The optical frequency of the probe is locked at around 0.7\,GHz below the $F = 4 \rightarrow F' = 5$ transition to minimize power broadening while still having appreciable (20\,mrad peak) polarization rotation.
A lock-in amplifier (SR830) demodulates the rotation signal with the local oscillator as a reference. 
The phase of the lock-in is chosen to make the \textsf{Y} trace completely dispersive and the \textsf{X} trace completely absorptive. 






To compare the performance, we measure polarimeter noise and magnetic resonance (MR) signal, as well as the response of the magnetometers to small steps in LO frequency. The former provides shot-noise-projected sensitivity while the latter is a direct measurement of the sensitivity of the experiment to small changes in the magnetic field.


We obtain the MR signal (Fig.~\ref{fig:Figure2}) by stepping the frequency of the local oscillator over the MR line and letting the lock-in output equilibrate at each frequency before acquiring the \textsf{X} and \textsf{Y} values. 
Then we fit the MR signal to a Lorenzian resonance and extract the amplitude and width. We measure the shot noise by repeating the scan with the pump beam blocked.
The atoms are not polarized in the rf-driven magnetometer without the pump, and the signal is dominated by the probe-laser shot noise equivalent to \shotnoise{} (black trace in the RF graph of Fig.~\ref{fig:Figure2}). 
In the LS magnetometer, however, the intensity-modulated light-shift beam creates a small amount of polarization (black trace in the LS graph of Fig.~\ref{fig:Figure2}). This establishes a Bell-Bloom type of synchronous pumping \cite{BellBloom} resonance due to residual photon scattering of the intensity-modulated LS beam.
Comparing the slope of the dispersive signal (blue) with the RMS noise without the pump results in a signal-to-noise of $1.2 \times 10^5$ which translates into the same projected sensitivity of \shotnoise{}. 

The ``parasitic'' resonance is 90$^{\circ}$ out of phase with the MR signals shown in Fig. 2 because direct synchronous excitation by the LS beam instantaneously creates polarization directed along \^{x} which precesses in the \^{x}-\^{y} plane. At the time of peak LS beam intensity, the orientation generated in this way is orthogonal to the probe beam path and causes no optical rotation of the probe. In the LS magnetometer, the peak intensity of the fictitious magnetic field causes the pumped atoms to be tipped from the \^{z} direction toward \^{y}, thus making the maximum optical rotation in phase with the LS beam intensity modulation. 
For stable laser parameters the additional resonance introduces a constant distortion, rather than noise, to the magnetometer signal. The amplitude of this resonance however depends on several parameters, e.g. LS beam detuning, intensity and polarization. If the magnetic field is determined via the zero crossing of the MR phase signal, parameter drifts of the LS laser cause systematic shifts of the observed MR frequency. It is the experience of the authors that these effects can be substantial under certain experimental conditions, and it seems plausible that this mechanism may play a role in the low-frequency stability of alkali-vapor magnetometers \cite{Ignacio}.

%

\begin{figure}[bth]
\centering
\includegraphics{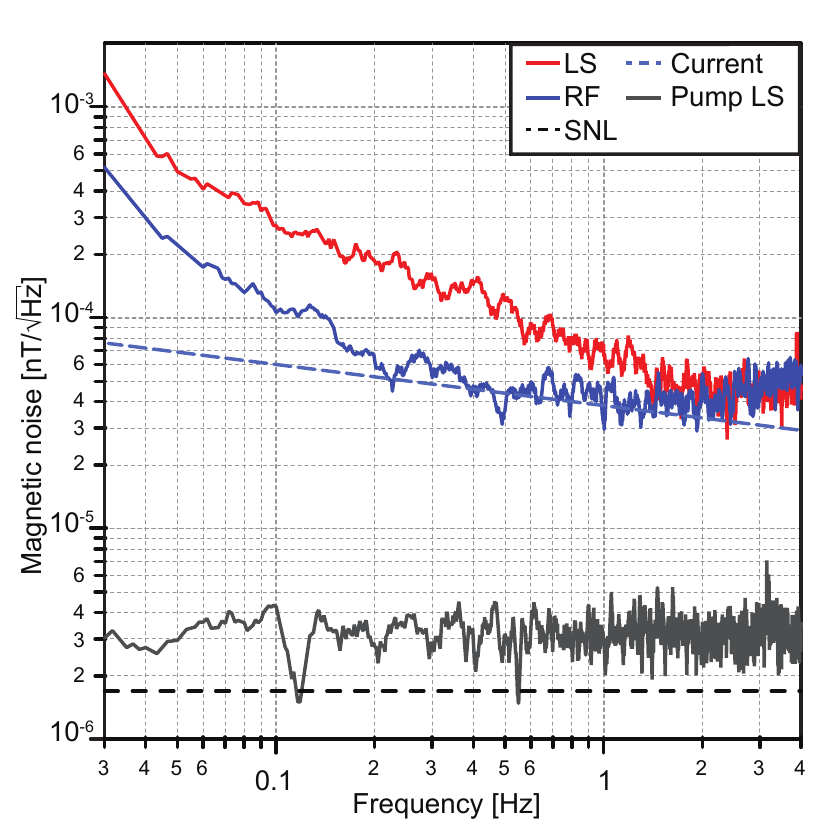}
\caption{Linear spectral density plot for the RF (dark blue) and LS (red) modes. The black dashed  line represents the shot noise limit (\shotnoise{}). The gray trace is the noise induced by the varying light shift caused by pump laser power fluctuations. (\pumpnoise{}). The blue dashed line is the magnetic field noise according to the specifications of the current source creating the leading field.}
\label{fig:LSD}
\end{figure}

\begin{figure}[bth]
\includegraphics{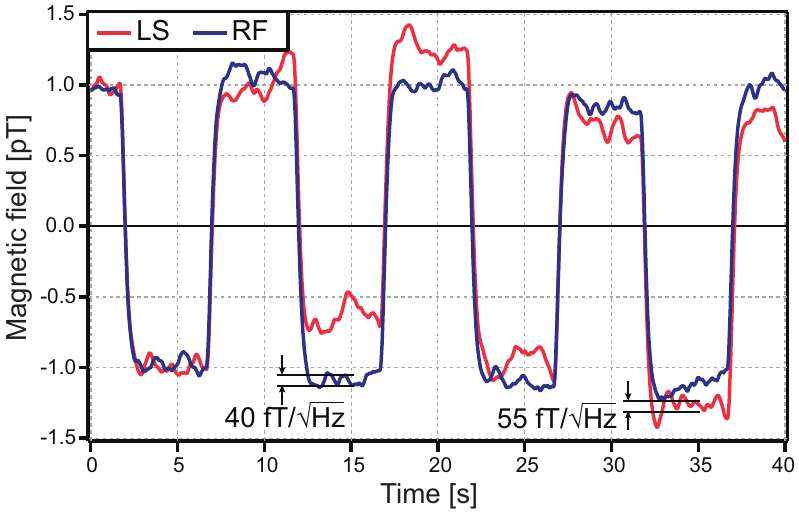}
\caption{Response of both magnetometers to steps in the LO frequency. The frequency generator was stepped by $\pm 3.5$\,mHz around the center of the magnetic resonance every 10\,s. The data were acquired with a noise bandwidth of 0.8~Hz due to the selected time constant of the lock-in amplifier.
}
\label{fig:steps}
\end{figure}
The ``parasitic'' synchronous pumping resonance observed in the LS configuration can also be used to measure the magnetic field quite effectively if the LS beam is tuned close to the absorption line of the atomic transition.
Optimization in this pumping mode results in projected sensitivity of \shotnoiseSync{} with 43\,$\mu$W average pump power and a 3\% duty cycle. For best performance in Bell-Bloom mode, the LS beam is tuned to address the $D2$ $F=3$ manifold. 
However, the presence of cycling transitions in the synchronous pumping scheme causes considerable MR line broadening (2.3\,Hz) compared to the LS magnetometer scheme (1.8\,Hz) which takes advantage of light narrowing. The difference could be even more drastic if an alkene cell with $\sim$77\,s relaxation time is utilized \cite{supercell}. 
For clarity, we focus in the analysis on the two (fictitious or real magnetic field) rf-driven magnetometers.

We compare the performance of those magnetometers by recording both, the linear spectral density (LSD) of the sensor signal (Fig. \ref{fig:LSD}) and the response to steps in the LO driving frequency (Fig. \ref{fig:steps}). 
To measure the LSDs, we positioned the frequency generator at the center of the MR and recorded the \textsf{Y} output of the lock-in. 
We observe the signal over time, convert it to an equivalent field shift and perform a Fourier transform of the result.
The comparison of both operation modes can be seen in Fig. \ref{fig:LSD}. 
Since the bandwidth of the magnetometer is constrained by its MR line width, we recorded the frequency response of the sensor and corrected the magnetic field noise accordingly. Both magnetometers exhibited a minimum noise floor of \labnoiseLSD{} at a frequency $\sim$ 2\,Hz, 
where the contribution of the laboratory magnetic noise is minimal. 


For the field step response measurement, we modulated the driving frequency around the Larmor resonance by $\pm$3.5\,mHz, effectively equivalent to a field change of $\pm$1\,pT. 
To estimate the noise, we observed the dispersive output of the lock-in and compared the step size to the RMS noise of the signal on a step level. The average RMS noise on the step was \labnoiseSTEP{} for the RF mode, in agreement with the LSD measurement. For the LS mode the RMS noise was slightly larger (\labnoiseSTEPLS{}) due to the increased  noise level at frequencies around 0.2\,Hz. 


In conclusion, we present an all-optical light shift magnetometer with projected sensitivity of \shotnoise{} and demonstrate its performance in a laboratory setup with a noise floor of \labnoiseLSD{} at 2\,Hz. 
The magnetometer employs an oscillating fictitious magnetic field created by an intensity-modulated circularly polarized laser beam to drive the precession in \cesium{}. 
We compare it to the same magnetometer driven by a real oscillating magnetic field and demonstrate similar performance. 
The light-shift magnetometer has several advantages compared to the conventional rf-driven magnetometer due to the absence of an actual rf magnetic field. It can be readily implemented in most current rf-driven magnetometer setups. 
This offers potential improvements in applications where the sensor density is of interest and arrays of magnetometers are in use, e.g., in the search for an electric dipole moment of the neutron.



%
%
%
%
%
%
%
\begin{acknowledgments}
A. W. was supported by a Marie Curie International Research Staff Exchange Scheme Fellowship within the 7th European Community Framework Programme. B. P. is supported by DFG Priority Program SPP1491, “Precision Measurements with Cold and Ultracold Neutrons.”
\end{acknowledgments}.
\bibliography{magnetometer_literature}

\end{document}